\documentclass[a4paper]{jpconf}
\usepackage{graphicx}
\begin{document}
\title{Meson spectroscopy with COMPASS}

\author{Frank Nerling \textnormal{\it on behalf of the COMPASS collaboration}}

\address{Physikalisches Institut, Albert-Ludwigs-Universit\"at Freiburg}

\ead{nerling@cern.ch}

\begin{abstract}
The COMPASS fixed-target experiment at CERN SPS is dedicated
to the study of hadron structure and dynamics. In the physics
programme using hadron beams, the focus is on the detection of
new states, in particular the search for $J^{PC}$ exotic states and
glueballs. After a short pilot run in 2004 (190 GeV/c negative
pion beam, lead target), we started our hadron spectroscopy
programme in 2008 by collecting an unprecedented statistics with
a negative hadron beam (190 GeV/c) on a liquid hydrogen target.
A similar amount of data with positive hadron beam (190 GeV/c)
has been taken in 2009, as well as some additional data with negative beam
on nuclear targets. The spectrometer features a large
angular acceptance and high momentum resolution and also good
coverage by electromagnetic calorimetry, crucial for the detection
of final states involving $\pi^0$ or $\eta$.
A first important result is the observation of a significant
$J^{PC}$ spin exotic signal consistent with the disputed $\pi_1(1600)$ in
the pilot run data. This result was recently published.
We present an overview of the status of various ongoing
analyses on the 2008/09 data.
\end{abstract}

\section{Introduction}
The existence of exotic states beyond the constituent quark model (CQM) has been speculated about almost since the introduction of 
colour~\cite{Jaffe:1976,Barnes:1983}. 
Due to the self-coupling of gluons via colour-charge, so-called hybrid mesons and glueball are allowed within Quantum Chromodynamics 
while they are beyond the CQM. Hybrid mesons are $q\bar{q}$ states with an admixture of gluons, and glueballs are states with no 
quark content, only consisting of (constituent) gluons. According to Lattice-QCD predictions \cite{cmcneile:2006}, the glueball candidate 
lowest in mass has scalar quantum numbers, $J^{PC}= 0^{++}$, and is expected to have a mass of $\sim 1.7\,{\rm GeV/c}^2$. A glueball 
candidate has been observed by the Crystal Barrel and the WA102 experiments, however, mixing with ordinary 
isoscalar mesons makes the interpretation difficult. 
Several light hybrids, on the other hand, are predicted to have exotic $J^{PC}$ quantum numbers and are thus promising candidates in the 
search for physics beyond the CQM.
The lowest mass hybrid candidate for example is predicted~\cite{Morningstar:2004} to have exotic quantum numbers of spin, parity and charge 
conjugation $J^{PC}=1^{-+}$ not attainable by ordinary $q\bar{q}$ states, and a mass between 1.3 and 2.2\,GeV/c$^2$.

Two experimentally observed $1^{-+}$ hybrid candidates in the light-quark sector have been reported in the past in different decay 
channels, the $\pi_1(1400)$ mainly seen in $\eta\pi$ decays, by e.g. E852~\cite{E852}, VES\cite{Beladidze:1993}, and Crystal 
Barrel~\cite{CB}, and the $\pi_1(1600)$, observed by both E852 and VES in the decay channels: $\rho\pi$~\cite{Adams:1998,Khokhlov:2000}, $\eta'\pi$~\cite{Beladidze:1993,Ivanov:2001}, $f_{1}\pi$~\cite{Kuhn:2004,Amelin:2005}, and $\omega\pi\pi$~\cite{Amelin:2005,Lu:2005}. In particular the resonant nature of the $\rho\pi$ decay channel of the $\pi_1(1600)$ observed in $3\pi$ final states is highly disputed~\cite{Amelin:2005,Dzierba:2006}. COMPASS has started to shed new light on the puzzle of spin-exotics by the observation of an $1^{-+}$ signal in the 2004 data, consistent with the famous $\pi_1(1600)$. It shows clean phase motions with respect to other waves, confirming the resonance 
nature~\cite{Alekseev:2009a}.            

\section{The COMPASS experiment}
The COMPASS two-stage spectrometer~\cite{compass:2007} at the CERN SPS features electromagnetic calorimetry in both stages. 
Photon detection in a wide angular range at high resolution is crucial for decay channels involving $\pi^{0}$, $\eta$ or $\eta'$,
\begin{figure}[tp!]
  \begin{minipage}[h]{.59\textwidth}
    \begin{center}
     \includegraphics[clip, trim= 45 60 55 80,width=1.0\linewidth]{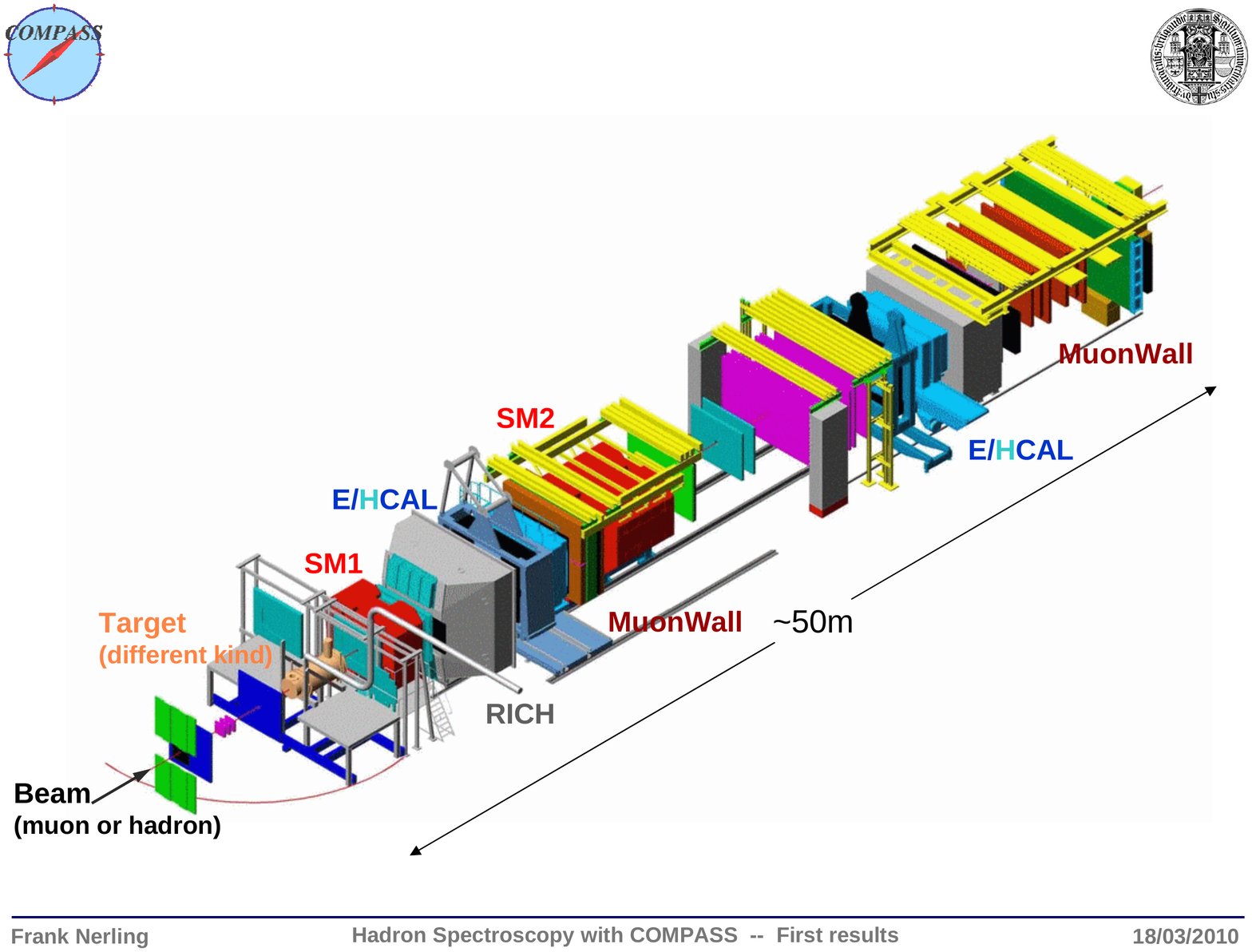}
    \end{center}
  \end{minipage}
  \hfill
  \begin{minipage}[h]{.39\textwidth}
    \begin{center}
     \includegraphics[clip, trim= 0 0 0 0,width=1.0\linewidth]{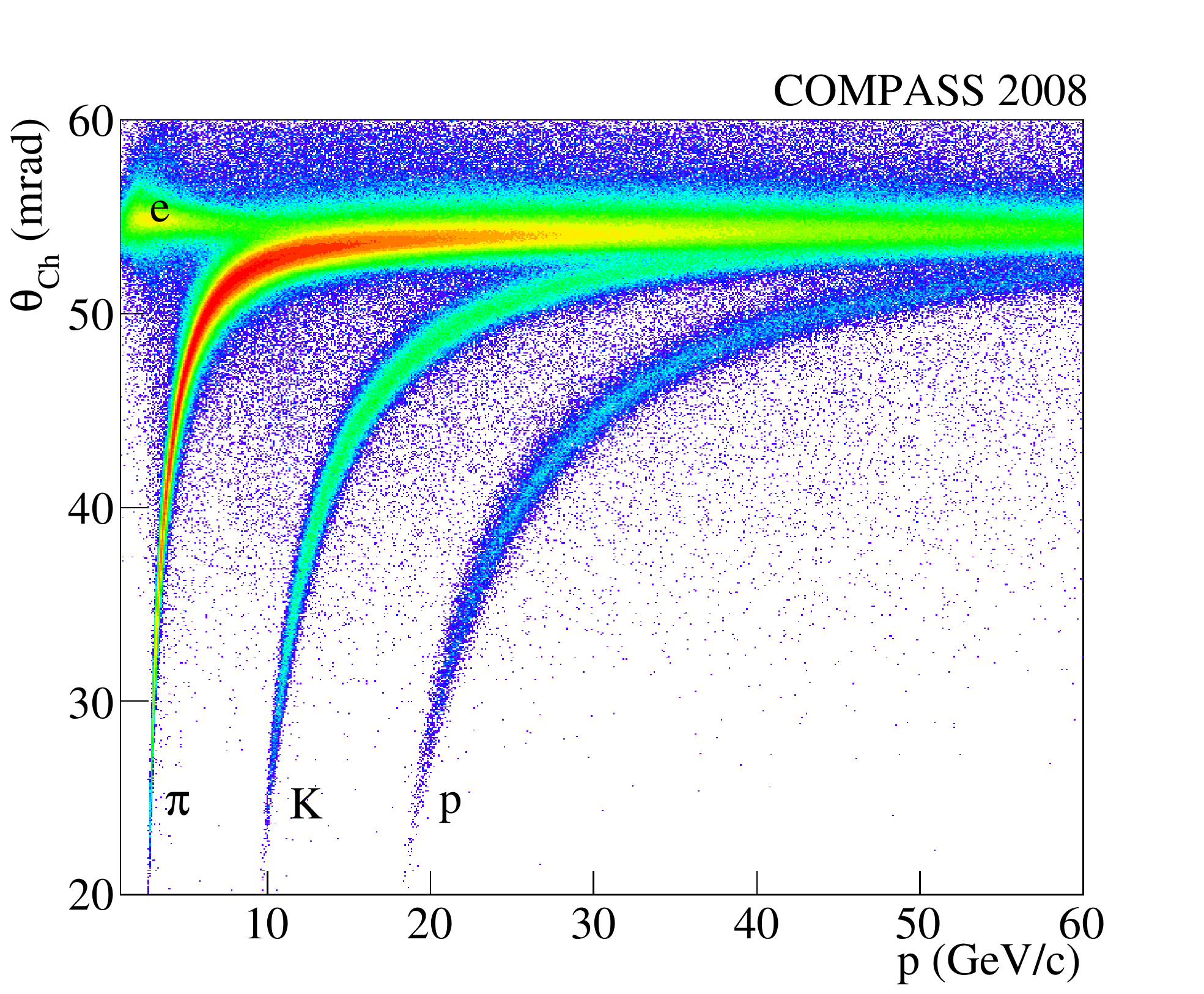}
    \end{center}
  \end{minipage}
      \caption{{\it (Left)} Sketch of the two-stage COMPASS spectrometer.
               {\it (Right)} Measured Cherenkov angle using 
	RICH-1 versus particle momentum. Three bands appear corresponding to the different masses of
	pions, kaons, and (anti-)protons; some additional contribution from $\delta$-electrons is present at low masses and angles.}
       \label{fig:diffrProd_Spectro} 
\end{figure}
\begin{figure}[bp!]
  \begin{minipage}[h]{.49\textwidth}
    \begin{center}
      \includegraphics[clip,trim= 20 0 0 0,width=0.75\linewidth,
       angle=0]{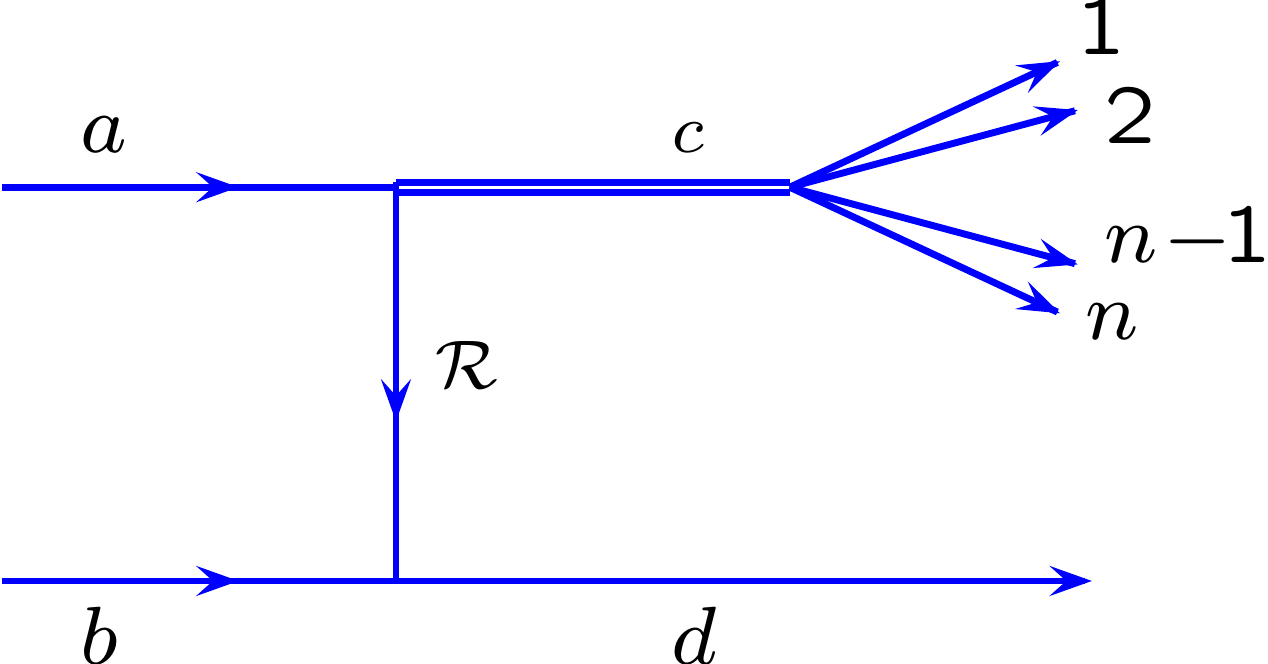}
    \end{center}
  \end{minipage}
  \hfill
  \begin{minipage}[h]{.49\textwidth}
    \begin{center}
      \includegraphics[clip,trim= 17 10 3 -10,width=0.75\linewidth,
     angle=0]{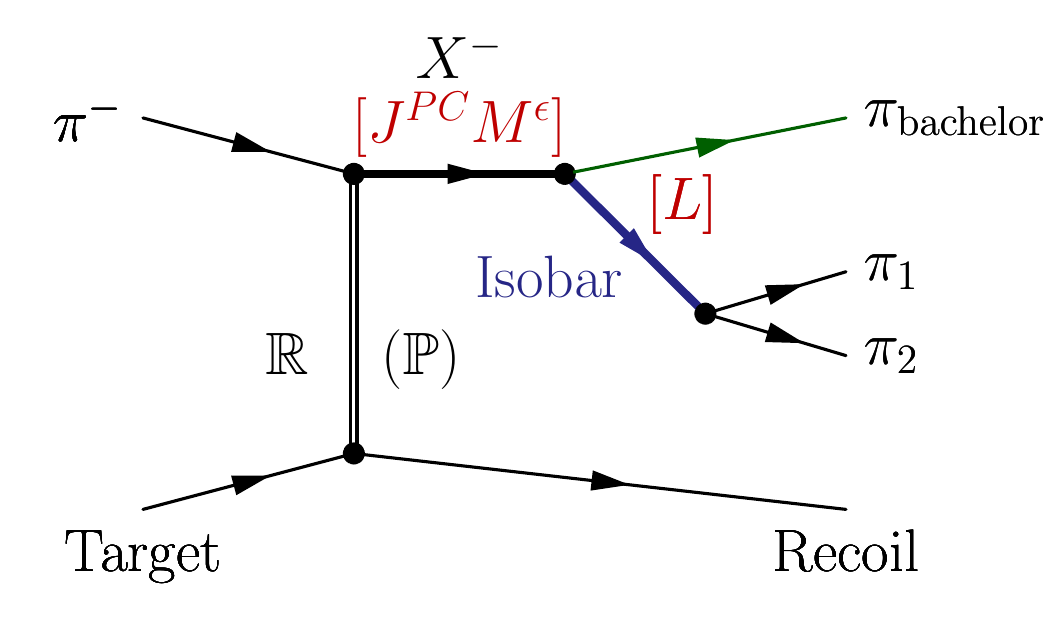}
    \end{center}
  \end{minipage}
      \caption{{\it (Left)} Diffractive meson production: The beam particle $a$ is excited, via $t$-channel Reggeon exchange, to a 
	resonance $c$ subsequently decaying into $n$ mesons, the target stays intact. {\it (Right)} Diffractive dissociation into 3$\pi$ 
	final states as described in the isobar model: The produced resonance $X^{-}$ with quantum numbers $J^{PC}M^\epsilon$ decays into 
	an isobar with spin $S$ and relative orbital angular momentum $L$ with respect to the $\pi_{\rm bachelor}$. The isobar subsequently 
	decays into two pions. At high energies, the Pomeron is the dominant Regge-trajectory.~~~~~~~~~~~~~~~~~~~~~~~~~~~~~~~~~~~~~~~~~~~~~~~~~~~~~~~~~~~~~~~~~~~~~~~~~~}
       \label{fig:diffrProd_Spectro} 
\end{figure}
The Ring Imaging Cherenkov (RICH) detector in the first stage allows for final state particle identification (PID). A good separation of pions from kaons enables the study of kaonic final states. 
Two Cherenkov Differential counters with Acromatic Ring focus (CEDAR) upstream of the target are used to identify the incoming beam particle.  
Not only production of strangeness with the pion beam can thus be studied but 
also kaon diffraction, tagging the kaon contribution in the negative hadron beam (96\,\% $\pi^{-}$, 3.5\,\% $K^{-}$, 
0.5\,\% $\bar{p}$). After a short pilot run in 2004 of 190 GeV/c $\pi^{-}$ beam on a lead target (Sec.\,\ref{subsec.2004}), 
COMPASS recorded data with 190\,GeV/c hadron beams in 2008/09, providing excellent opportunity for simultaneous observation of 
new states in various decay modes within the same experiment (Sec.\,\ref{subsec.2008}, \ref{subsec.2008b}). 
Moreover, the data contain subsets with different beam projectiles ($\pi^{\pm},K^{\pm},p$) and targets (H$_2$, Ni, W, and Pb), allowing for systematic studies not only of diffractive and central production but also Primakoff reactions~\cite{grabmueller:2010} and baryon spectroscopy~\cite{austregesilo:2010}.  

\subsection{Observation of a $J^{PC} = 1^{-+}$ exotic resonance -- 2004 data}
\label{subsec.2004}
The quantum numbers spin $J$, parity $P$ and $C$-parity of the produced resonance $X^{-}$, together with 
the spin projection given by $M$ and $\epsilon$ (reflectivity), define a partial wave $J^{PC}M^\epsilon[isobar]L$.
The partial wave analysis (PWA) is based on the isobar model, see Fig.\ref{fig:diffrProd_Spectro} {\it (right)}. 
The resonance $X^{-}$ decays via an intermediate di-pion resonance (the isobar), accompanied by a so-called bachelor pion.
The PWA method consists of two steps. First, a mass-independent fit is performed on the data binned into 40\,MeV/c$^2$ wide 
mass intervals, no assumption on the resonance structure of the $3\pi$ system is made at this level. 
A total set of 42 waves including a flat background wave is fitted to the data using an extended maximum likelihood method, which 
comprises acceptance corrections. 
Subsequently, the mass-dependent fit is applied to the six main waves out of the result from the first step, and uses a $\chi^2$ 
minimisation. 
The mass dependence is parameterised by relativistic Breit-Wigners (BW) and coherent background, if present. 
The employed parameterisation of the spin density matrix has a rank of two, accounting for spin-flip and spin-non-flip amplitudes 
at the baryon vertex. 
\begin{figure}[tp!]
\vspace{-0.6cm}
  \begin{minipage}[h]{.49\textwidth}
    \begin{center}
      \includegraphics[clip,trim= 3 4 22 5,width=0.9\linewidth,
       angle=0]{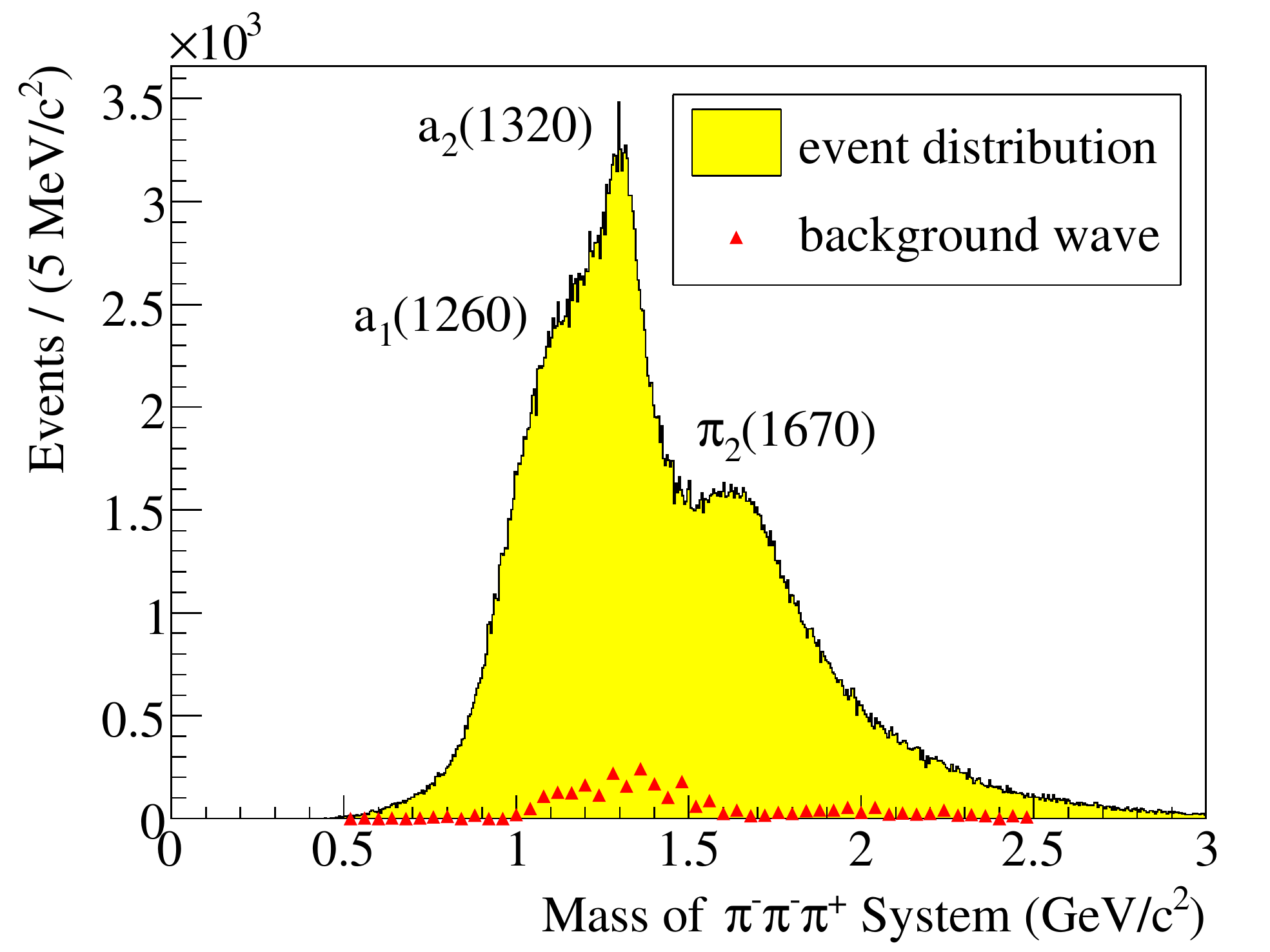}
    \end{center}
  \end{minipage}
  \hfill
  \begin{minipage}[h]{.49\textwidth}
    \begin{center}
      \includegraphics[clip,trim= 24 15 10 360,width=0.95\linewidth,
     angle=0]{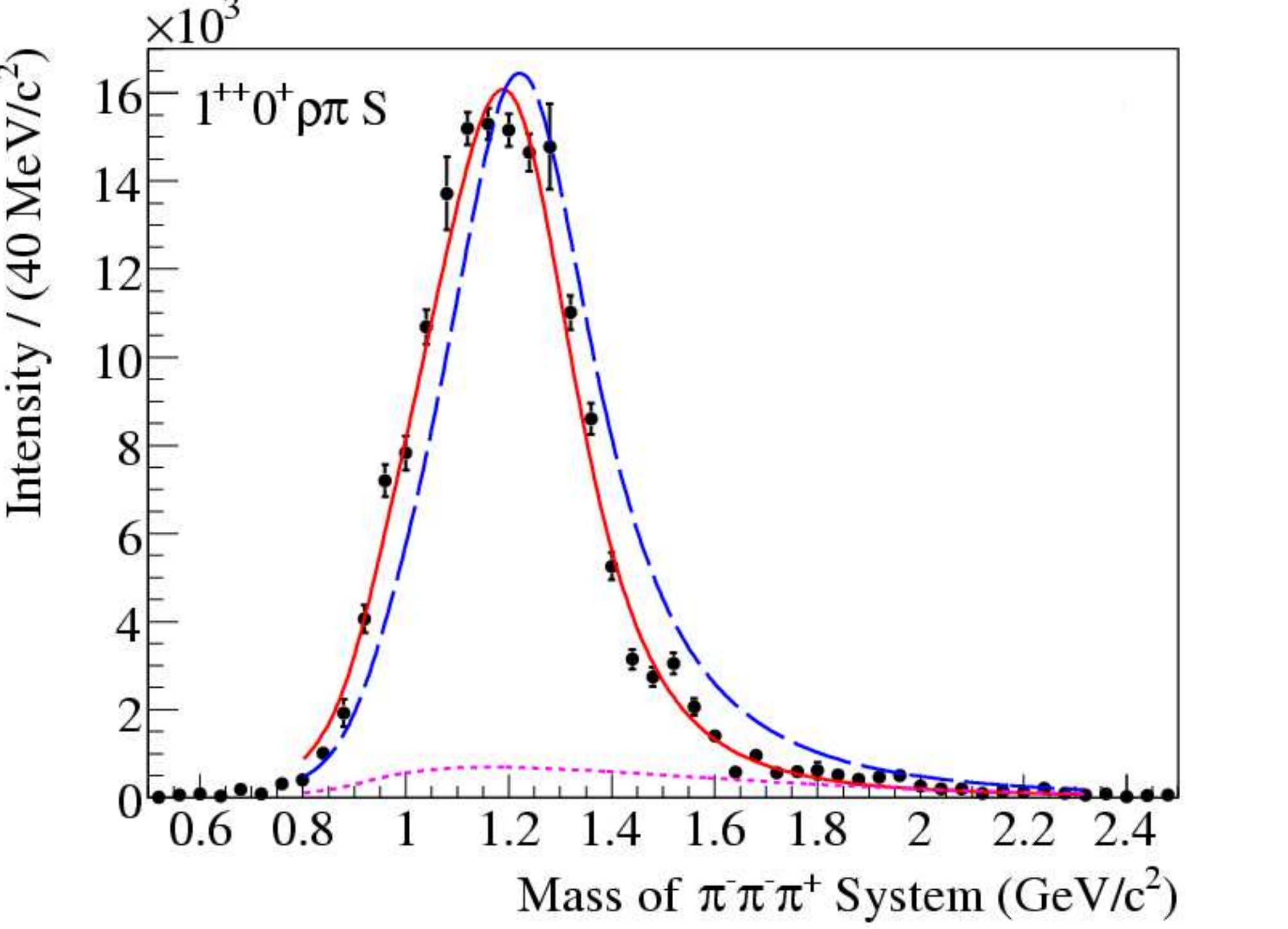}
    \end{center}
  \end{minipage}
\begin{minipage}[h]{.49\textwidth}
    \begin{center}
      \includegraphics[clip,trim= 4 15 30 360,width=0.95\linewidth,
	angle=0]{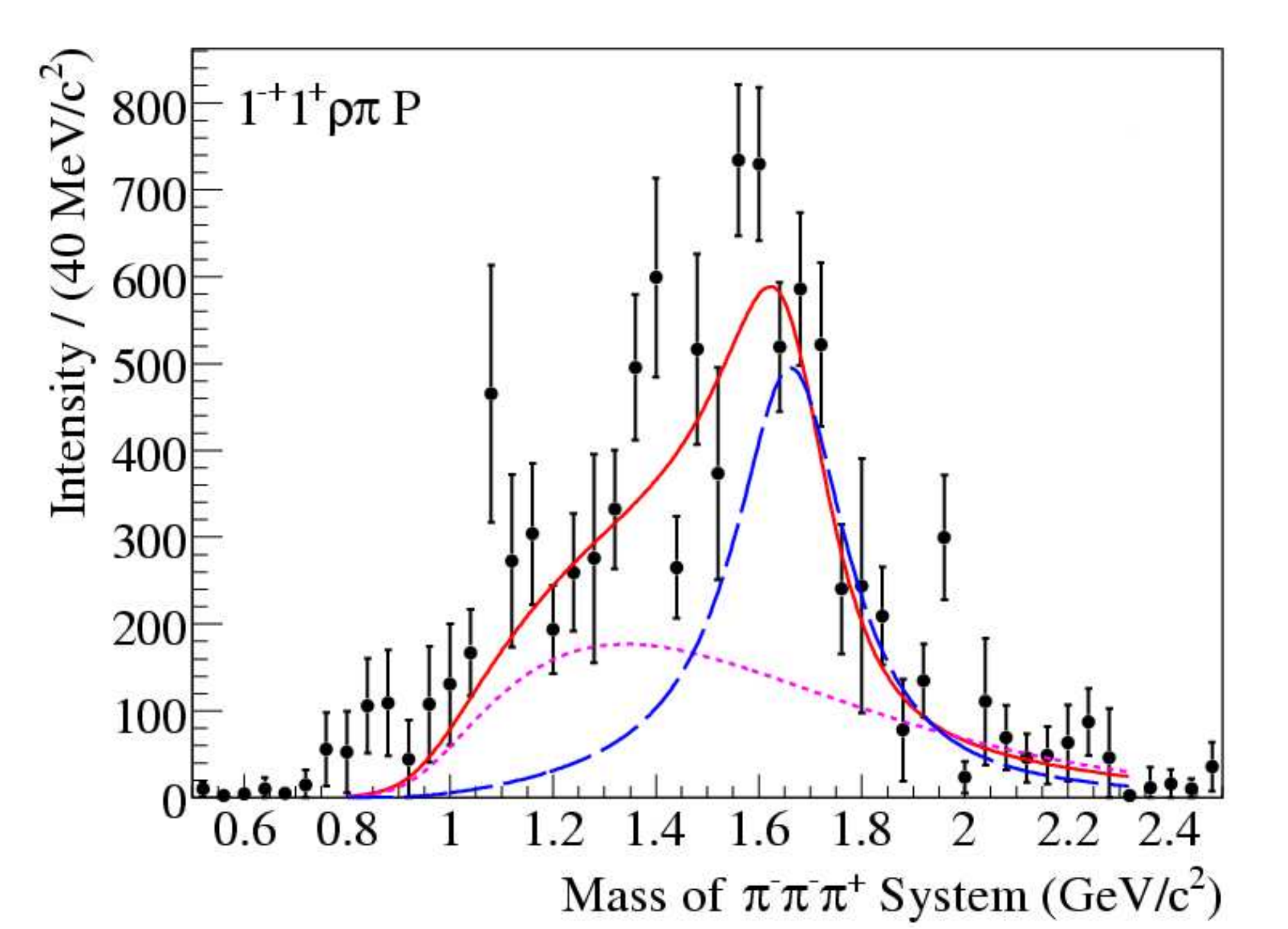}
    \end{center}
  \end{minipage}
  \hfill
  \begin{minipage}[h]{.49\textwidth}
    \begin{center}
      \includegraphics[clip,trim= 24 15 10 360,width=0.95\linewidth,
     angle=0]{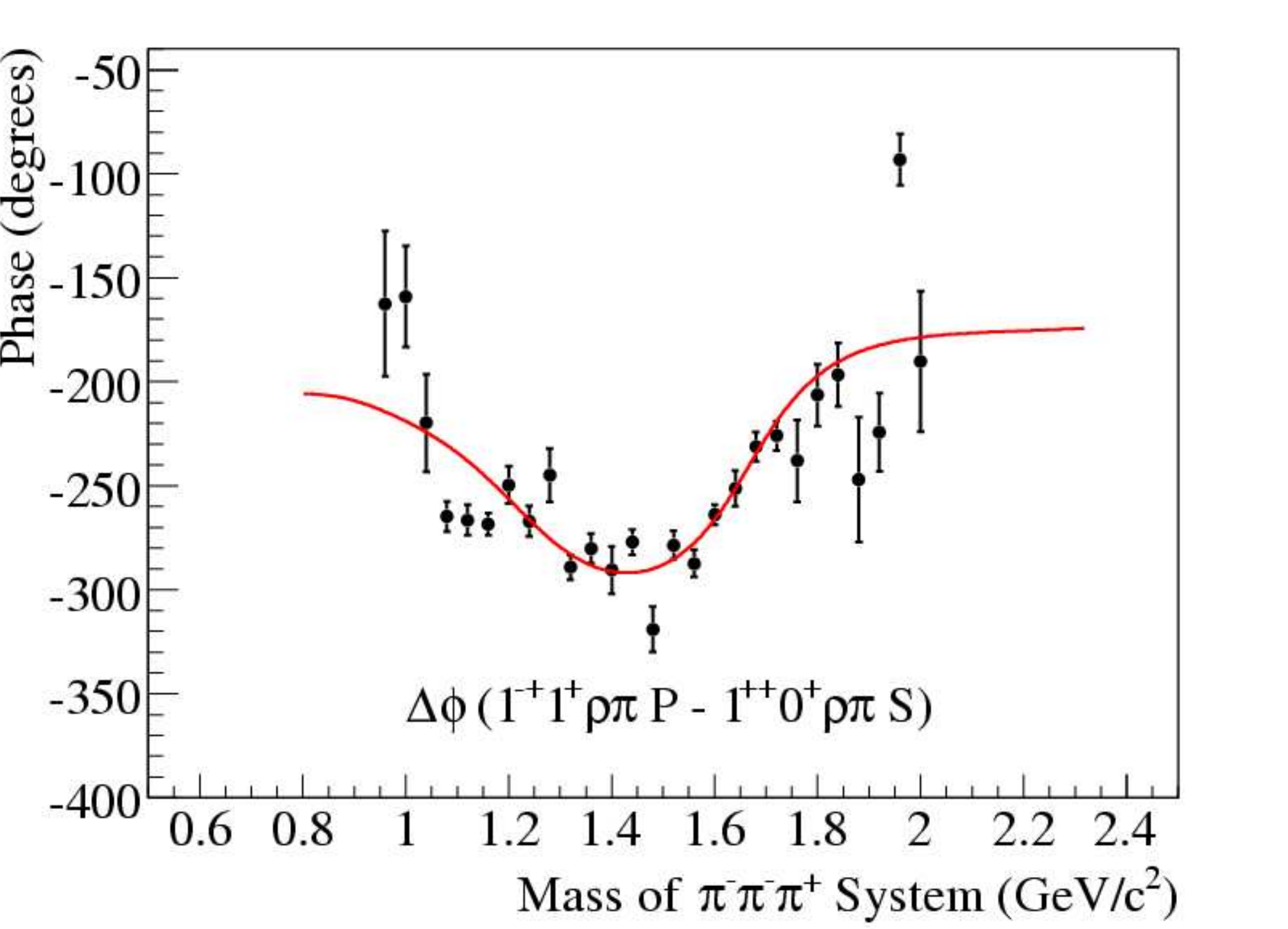}
    \end{center}
  \end{minipage}
      \caption{Top: {\it (Left)} $3\pi$ invariant mass, the most prominent resonances are indicated. {\it (Right)} Fitted intensity of 
	the exotic $1^{\rm -+} 1^{\rm +}[\rho \pi] P$ wave. Bottom: {\it (Left)} PWA fits for the $a_{\rm 1}(1260)$: Intensity of $1^{\rm ++} 0^{\rm +}[\rho^{-} \pi] S$ wave. {\it (Right)} Phase motion of the exotic $1^{\rm -+}1^{\rm +}$ versus $1^{\rm ++} 0^{\rm +}$ wave.}
      \label{fig:PWA2004}
\end{figure}
Fig.\,\ref{fig:PWA2004} shows the intensity of the $1^{++}$ wave with the well-established $a_1(1260)$ and that of the spin-exotic $1^{-+}$ wave 
as well as the phase difference $\Delta\Phi$ between the two, for the fits of all six waves, see \cite{Alekseev:2009a}. The black data points represent 
the mass-independent fit, whereas the mass-independent one is overlayed as solid line, the separation into background 
(dotted) and BW (dashed) is plotted where applicable. Especially the resonant nature of the exotic $1^{-+}1^{+}[\rho\pi] P$ wave is questioned in previous observations, whereas our data shows a clear and rapid phase motion.
Our result of a mass of $1660\pm 10^{+0}_{-64}$\,MeV/c$^2$ and a width of $269\pm 21^{+42}_{-64}$\,MeV/c$^2$ is consistent with the 
famous $\pi_1(1600)$~\cite{Alekseev:2009a} already reported in the past but still controversially discussed.

\subsection{First comparison of neutral versus charged mode  -- 2008 data {\small (negative beam, H$_2$ target)}}
\label{subsec.2008}
An important cross-check of all analyses is the test for isospin symmetry in the observed spectra.
\begin{figure}[tp!]
  \begin{minipage}[h]{.49\textwidth}
    \begin{center}
      \includegraphics[clip,trim= 3 4 22 5,width=0.9\linewidth,
       angle=0]{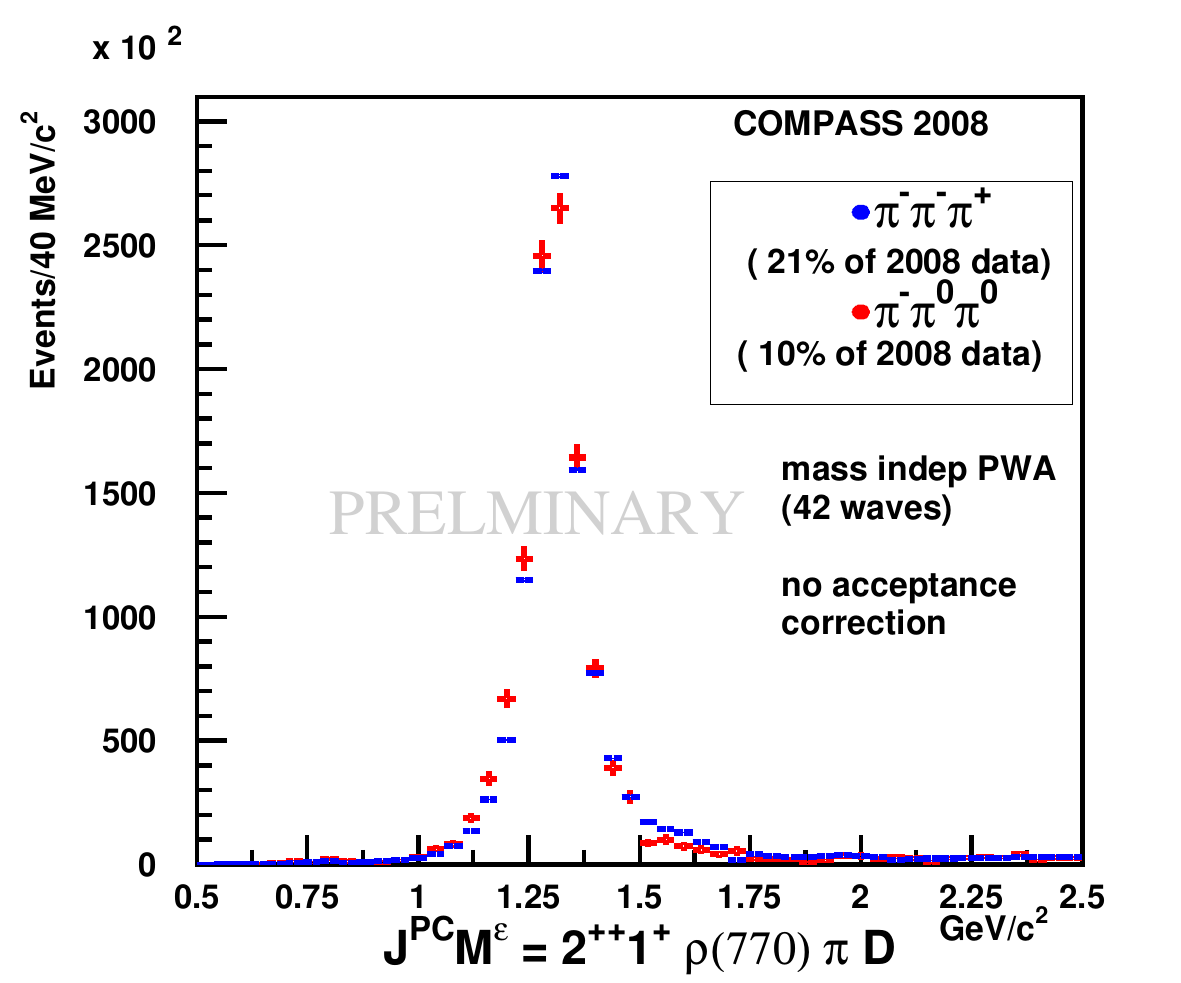}
    \end{center}
  \end{minipage}
  \hfill
  \begin{minipage}[h]{.49\textwidth}
    \begin{center}
      \includegraphics[clip,trim= 3 4 22 5,width=0.9\linewidth,
     angle=0]{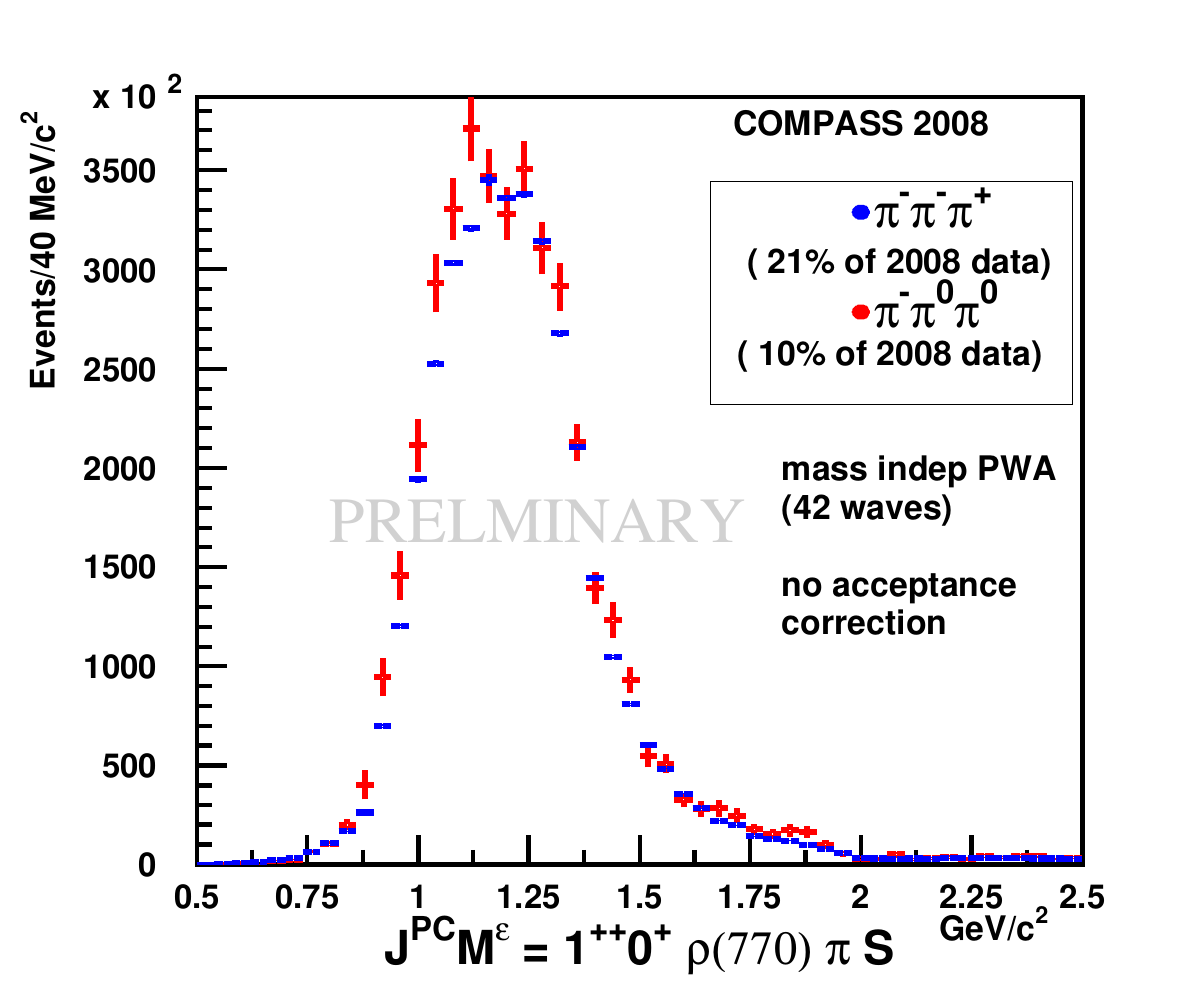}
    \end{center}
  \end{minipage}
  \begin{minipage}[h]{.49\textwidth}
    \begin{center}
      \includegraphics[clip,trim= 3 4 22 5,width=0.9\linewidth,
	angle=0]{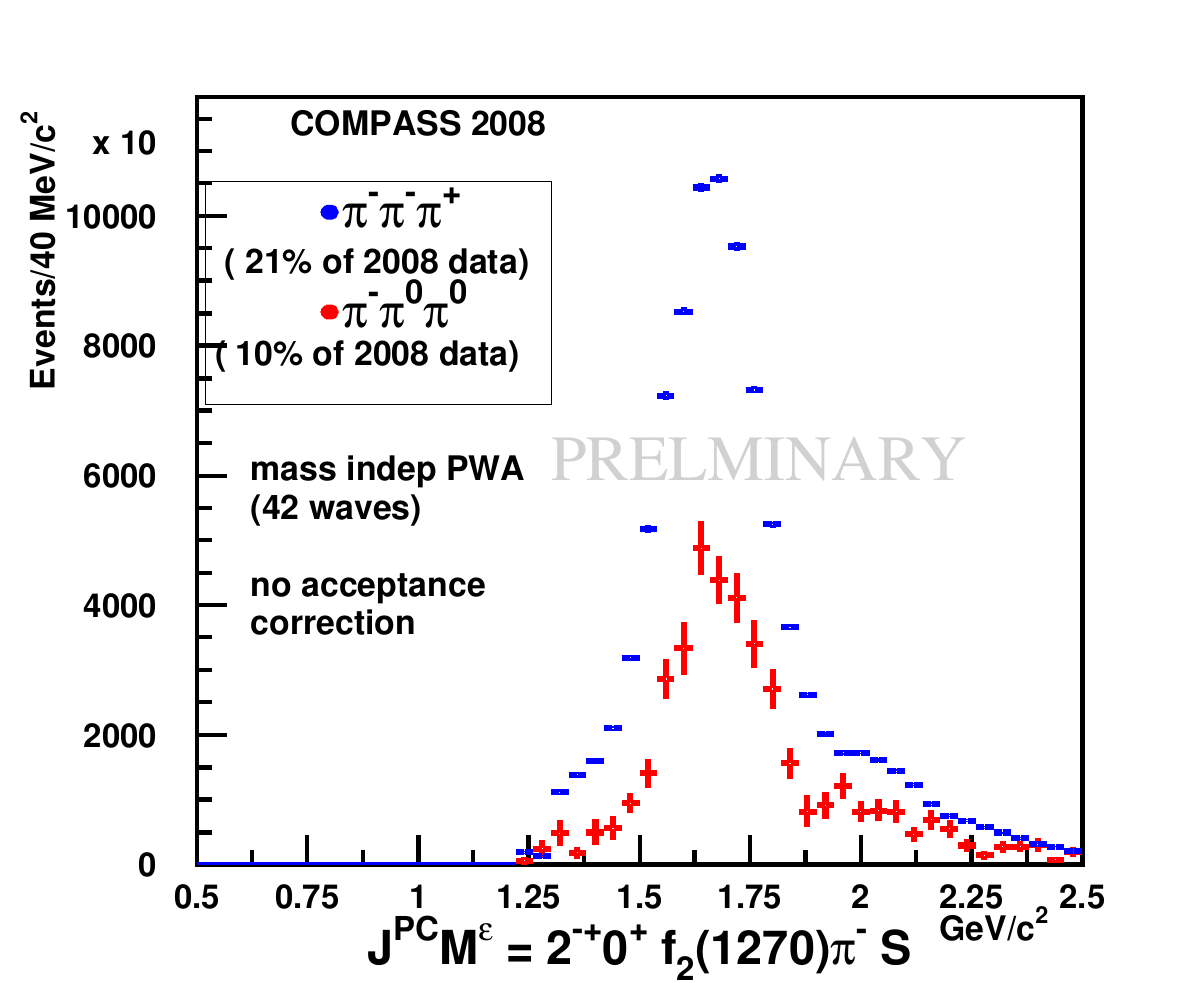}
    \end{center}
  \end{minipage}
  \hfill
  \begin{minipage}[h]{.49\textwidth}
    \begin{center}
      \includegraphics[clip,trim= 3 4 22 5,width=0.9\linewidth,
     angle=0]{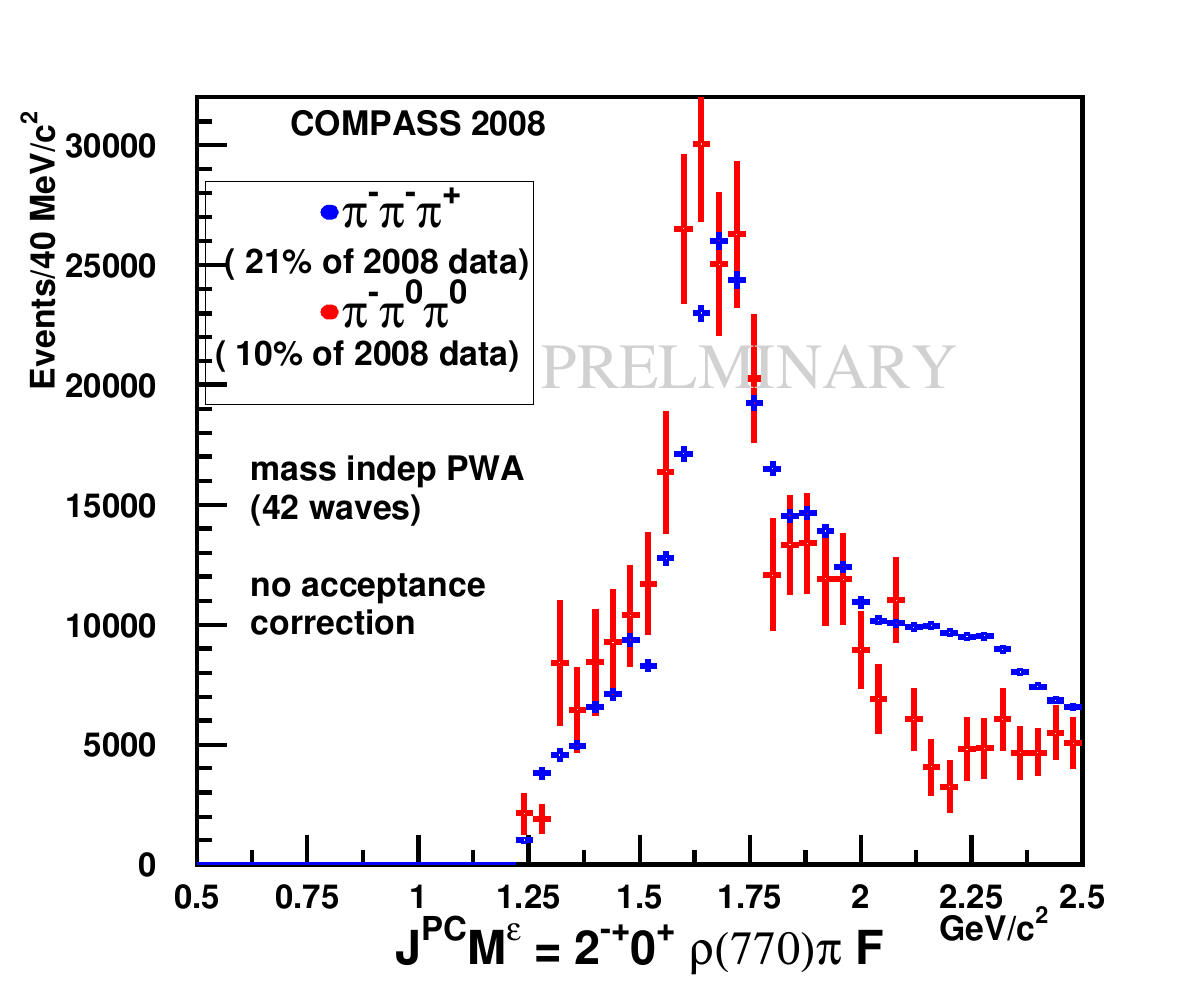}
    \end{center}
  \end{minipage}
      \caption{Comparison of PWA intensities of main waves for neutral vs. charged mode. \newline 
       {\it (Top/Left)} Intensities of the $a_{\rm 2}$ ($2^{\rm ++}1^{\rm +}$ going into $\rho^{-} \pi$ D wave) 
       used for normalisation of charged to neutral mode. Top/Right: ($a_{\rm 1}$) $1^{\rm ++} 0^{\rm +}$ 
       into $\rho^{-} \pi$ D wave. 
       {\it (Bottom/Left)} ($\pi_{\rm 2}$) $2^{\rm -+} 0^{\rm +}$ into $f_{\rm 2}$(1270) $\pi$ S wave. 
       {\it (Bottom/Right)} ($\pi_{\rm 2}$) $2^{\rm -+} 0^{\rm +}$ into $\rho^{-} \pi$ F wave.}
      \label{fig:PWA2008}
\end{figure}
The $\rho\pi$ decay channel of the $\pi_1(1600)$ for example, can be studied in two modes of 3$\pi$ 
final states, $\pi^{-}\pi^{+}\pi^{-}$ (charged) and $\pi^{-}\pi^{0}\pi^{0}$ (neutral), respectively.
The relative contribution should follow isospin conservation, depending on the underlying isobars, as
it is shown in Fig.\,\ref{fig:PWA2008}. 
A first partial-wave analysis (PWA) of main waves 
in diffractively produced 3$\pi$ events has been performed for both modes, applying the same model 
as for the 2004 result, for details see~\cite{nerling:2009}. 
The wave intensities shown are normalised in the same way to the well-known $\rho\pi$ decay 
of the $a_2(1320)$ to compensate for the different statistics analysed, thus making them comparable. We find similar intensities for the $\rho\pi$ decay, whereas a suppression factor of two is observed for 
the wave decaying into $f_2\pi$ as expected due to the Clebsch-Gordon coefficients.

\clearpage
Further ongoing analyses of neutral channels cover $\pi^{-}\eta$ and $\pi^{-}\eta\eta$ final states (search for the 
$\pi_1(1400)$ and lightest $0^{++}$ glueball candidate) as well as $\pi^{-}\pi^{-}\pi^{+}\pi^{0}$,
$\pi^{-}\pi^{-}\pi^{+}\eta$ and $\pi^{-}\pi^{-}\pi^{+}\pi^{0}\pi^{0}$ final states (accessible isobars: 
$f_1, b_1, \eta, \eta', \omega$). For all these channels, COMPASS has recorded significantly higher statistics 
with respect to previous experiments, covering all spin-exotic meson decay channels in the light quark sector 
reported in the past.     
\subsection{First glimpse on kaonic final states  -- 2008 data {\small (negative beam, H$_2$ target)}}
\label{subsec.2008b}

\begin{figure}[tp!]
  \begin{minipage}[h]{.49\textwidth}
    \begin{center}
      \includegraphics[clip,trim= 10 5 25 15,width=1.0\linewidth,
	angle=0]{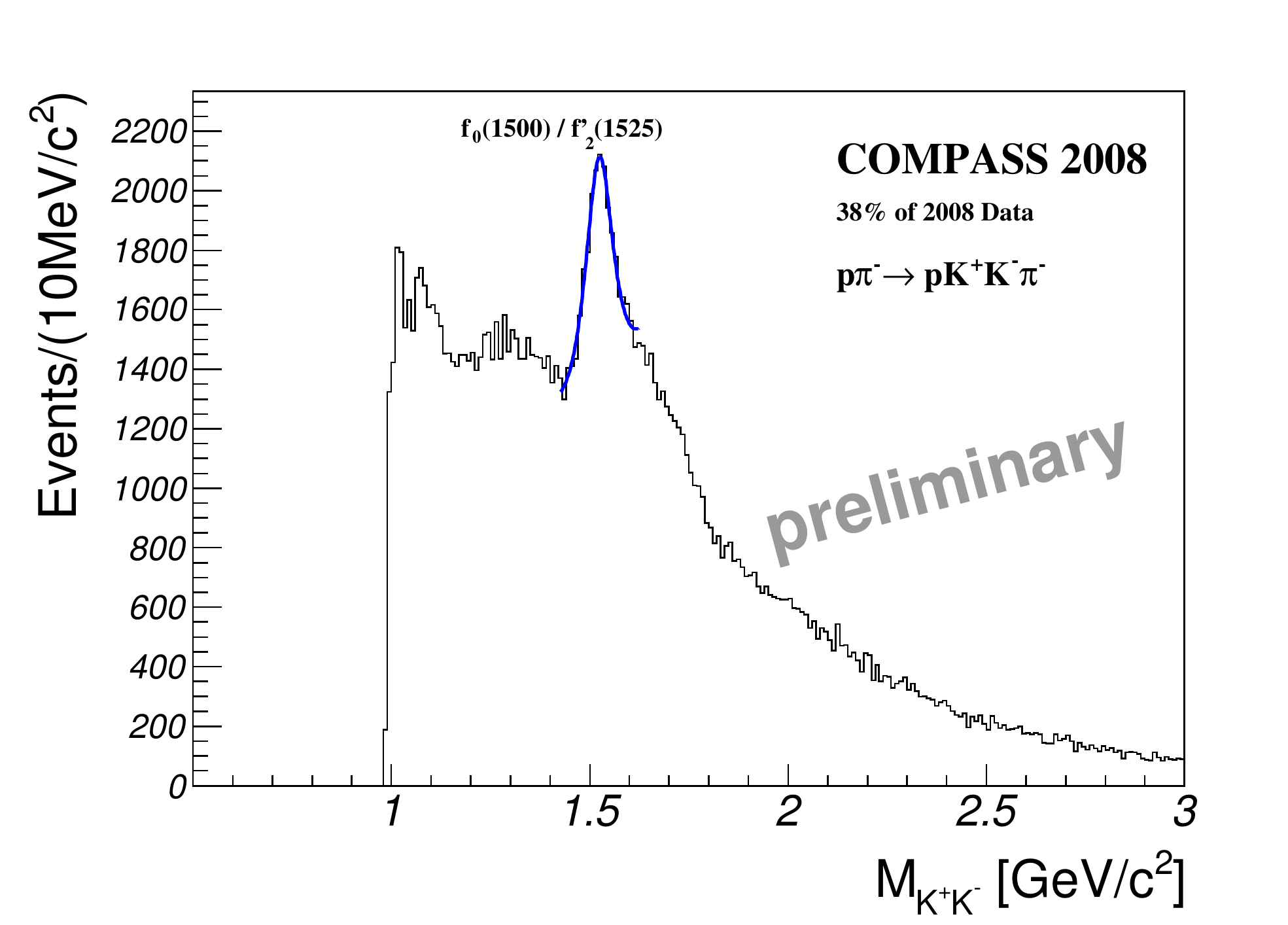}
    \end{center}
  \end{minipage}
  \hfill
  \begin{minipage}[h]{.49\textwidth}
    \begin{center}
       \includegraphics[clip,trim= 10 0 25 20,width=1.0\linewidth,
     angle=0]{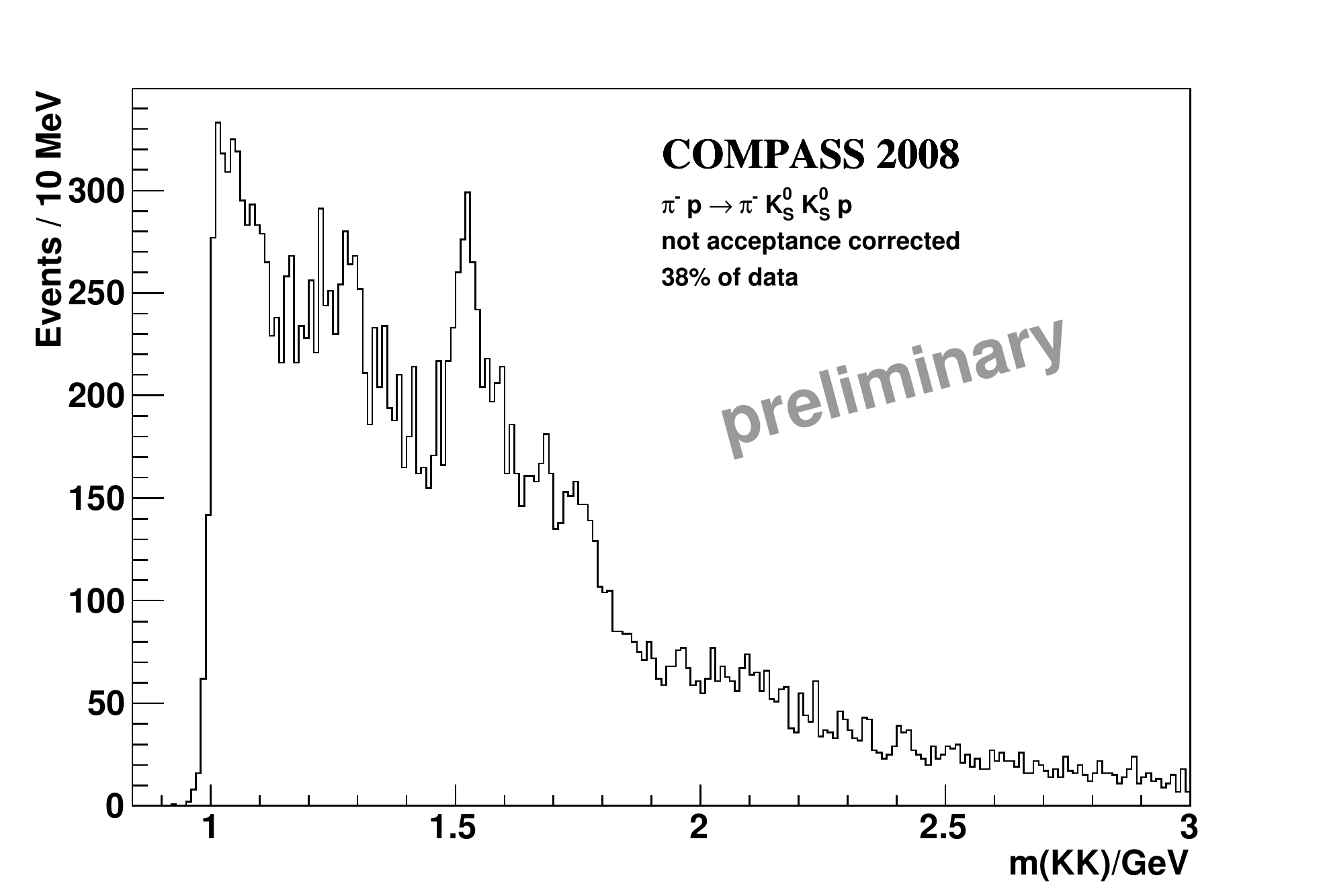}
    \end{center}
  \end{minipage}
      \caption{Invariant mass spectra of the diffractively produced $(K\bar{K}\pi)^{-}$ systems with the $\pi^{-}$ beam on the liquid hydrogen target: 
	{\it (Left)} $K^+K^-$ (with $p_{K^-} \le 30$\,GeV/c) {\it (Right)} $K^0_sK^0_s$.}
      \label{fig:Kaons}
\end{figure}

Final states including strange particles are interesting for both, glueball search in central production 
as well as diffractively produced hybrids. Fig.\,\ref{fig:Kaons} shows the $K\bar{K}$ subsystems 
out of the $(K\bar{K}\pi)^{-}$ system, again for two different modes: $K^{+}K^{-}\pi^{-}$ (charged mode) and 
$K^{0}_s K^{0}_s\pi^{-}$ (neutral mode) final states, respectively. 
In both cases, the spectra show a clear structure around the expected $f_0(1500)$. The CEDARs were used to 
anti-tag the kaons in the beam, and the final state kaons are identified using the RICH detector and the well resolved 
$V^{0}$ secondary vertex, respectively; for details on the event selections, see~\cite{tobi:2009}. 
\begin{figure}[bp!]
  \begin{minipage}[h]{.49\textwidth}
    \begin{center}
      \includegraphics[clip,trim= 10 5 25 15,width=1.0\linewidth,
	angle=0]{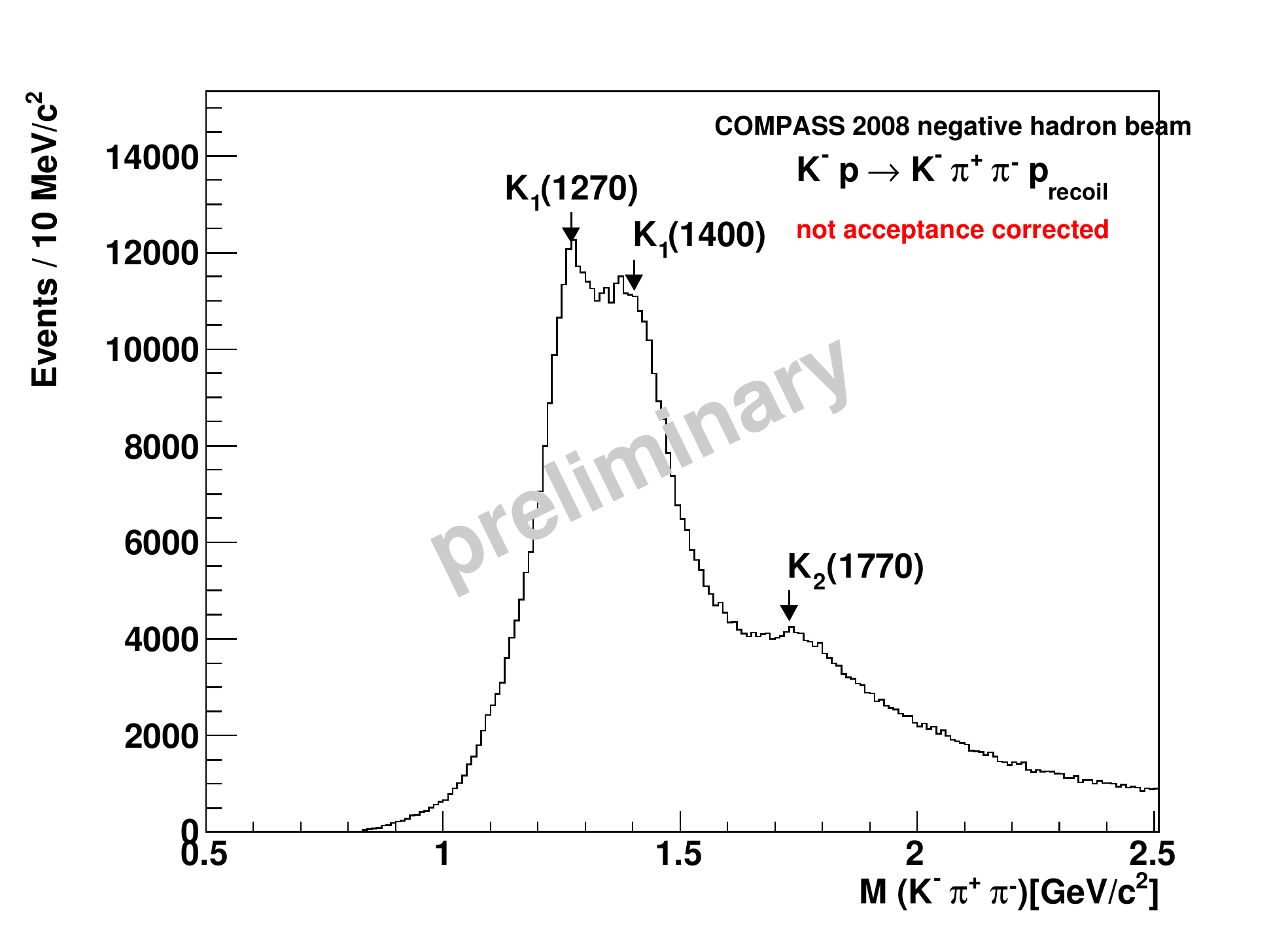}
    \end{center}
  \end{minipage}
  \hfill
  \begin{minipage}[h]{.49\textwidth}
    \begin{center}
      \includegraphics[clip,trim= 10 0 25 20,width=1.0\linewidth,
     angle=0]{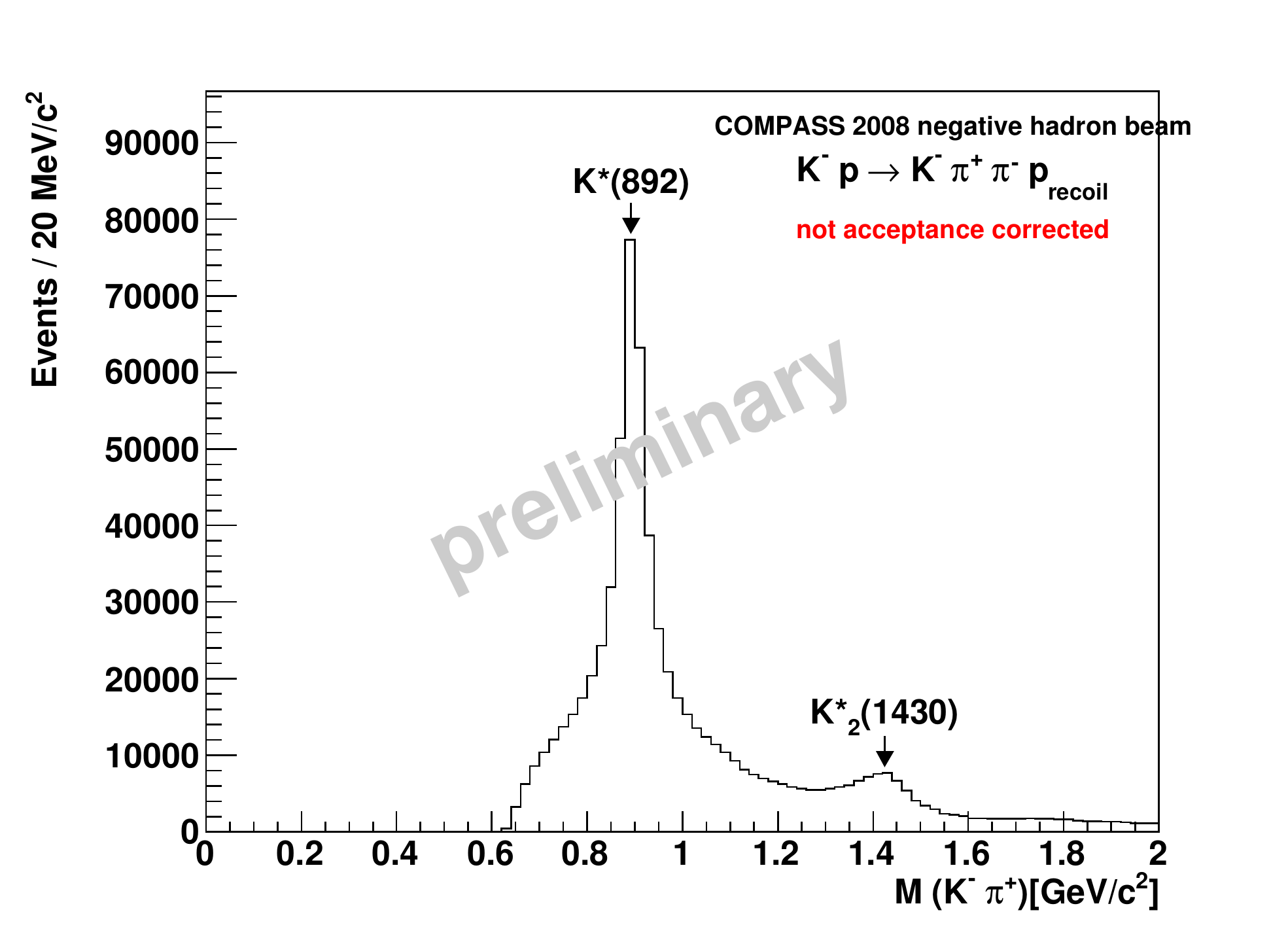}
    \end{center}
  \end{minipage}
      \caption{Kaon diffraction into $K^{-}\pi^{+}\pi^{-}$ final states:
	{\it (Left)} Invariant mass of the total diffractively produced system. 
	{\it (Right)} Invariant mass of the subsystem $K^{-}\pi^{+}$.}
      \label{fig:Kaons_b}
\end{figure}

Further ongoing analyses of kaonic final states cover the complementary $(K\bar{K}\pi)^{0}$ system diffractively produced with the $\pi^{-}$ 
beam as well as kaon diffraction into $K^{-}\pi^{+}\pi^{-}$ final states using the beam kaons. In both cases, the RICH detector is used for 
final state PID and the CEDARs for identifying the beam particle. 
Fig.\,\ref{fig:Kaons_b} shows mass spectra for the study of kaon diffraction into $K^{-}\pi^{+}\pi^{-}$ final 
states. The total mass spectrum and the $K^{-}\pi^{+}$ show the prominent, well-known resonances as expected, similar
as observed by WA03. Our statistics collected in merely 2008 exceeds the one from WA03 by a factor of two, for details, 
see \cite{promme:2009} and references therein.
The PWA of all kaonic final states mentioned in this paper are under preparation.
\section{Summary \& conclusions}
COMPASS has taken data with high-intensity, negatively as well as positively charged hadron beams 
($\pi^{\pm},K^{\pm},p$) on nuclear and liquid hydrogen targets. 
The data sample newly taken in 2008/09 exceeds the world data by a factor of 10-100, depending on 
the given final state. The COMPASS data sets allow to address open issues in light-mesons spectroscopy 
at good accuracy, even extending the region to higher masses beyond 2\,GeV/c$^2$.   

\section*{Acknowledgements}
This work is supported by the BMBF (Germany), particularly the ``Nutzungsinitiative CERN''.

\end{document}